\documentclass[aps,prd,twocolumn,superscriptaddress]{revtex4}

\usepackage[dvips]{graphicx}
\newcommand{\Dslash}{\slash\hspace{-2.5mm}D}

\begin{document}

% Definition of title page:
\title{A non-perturbative study of the action parameters for 
       anisotropic-lattice quarks}

\author{Justin Foley}
\author{Alan \'O Cais}
\author{Mike Peardon}
\author{Sin\'ead M. Ryan}
\affiliation{School of Mathematics, Trinity College, Dublin~2, Ireland}
\collaboration{TrinLat Collaboration}

\date{\today}
\preprint{TrinLatXXX}

\begin{abstract}
A quark action designed for highly anisotropic lattice simulations is discussed. The
mass-dependence of the parameters in the action is studied and the results are presented. 
Applications of this action in studies of heavy quark quantities are described and results 
are presented from simulations at an anisotropy of six, for a range of quark masses from
strange to bottom. 
\end{abstract}

\maketitle

%%%%%%%%%%%%%%%%%%%%%% 
\section{Introduction}
\label{sec:intro}
%%%%%%%%%%%%%%%%%%%%%%
%
The anisotropic lattice has proved an invaluable tool for simulations of a variety of physical
quantities. The precision calculation of the glueball spectrum was an early application of 
the approach~\cite{Morningstar:1999rf} and it was recognised that anisotropic actions may also
be advantageous in heavy quark physics calculations~\cite{Alford:1997nx}. 
Correlators of heavy particles such as glueballs and hadrons with a charm or bottom quark have a signal which 
decays rapidly. Monte Carlo estimates of these correlation functions can be noisy, making it difficult to 
resolve a plateau over a convincing range of lattice time steps. 
Increasing the number of timeslices for which the effective mass of a particle has reached
a plateau solves this problem and also decreases the statistical error in the fitted
mass. Since this value may be used as an input to determine many physical parameters this
decrease is very beneficial. 

Secondly, improved precision in effective mass fits means that momentum-dependent errors
of ${\cal O}(ap)$ can be disentangled from other discretisation effects and larger particle momenta may be
considered. This is particularly relevant for the determination of semileptonic decay form
factors where the overlap of momentum regions accessible to experiments and lattice
calculations is currently very small. Typically, experiments have more events with
daughter particle momentum at or above 1 GeV. This is also the region where large
momentum-dependent errors are expected in lattice calculations. The form factors of decays like
$B\rightarrow\pi\ell\nu$ and $B\rightarrow K^\ast\gamma$ are inputs to determinations of
CKM parameters so that increased precision in lattice calculations can lead to tighter 
constraints on the Standard Model. This has motivated a study of 2+2 anisotropic lattices where the 
temporal and one spatial direction are made fine and all momentum is injected along
this fine spatial axis. Details of the progress to date in this work are in
Refs.~\cite{Burgio:2003nk} and~\cite{Burgio:2003gx}. 
The 2+2 formulation has also proved useful for a precision determination of the static
interquark potential over large separations, which is described in Ref.~\cite{Burgio:2003nk}. 
In this paper we consider a 3+1 anisotropic action. The temporal lattice spacing, $a_t$ is
made fine relative to the spatial spacing, $a_s$. The action is designed with simulations at
large anisotropies in mind. To simulate a bottom quark with a relativistic action requires a 
lattice spacing of less than $0.04$ fm which is prohibitively expensive on an isotropic
lattice where the simulation cost scales at least as ${\cal O}(a^{-4})$. The anisotropic lattice offers the 
possibility of relativistic heavy-quark physics using reasonably modest computing resources. 
In the rest frame of a hadron with a heavy constituent, the quark four-momentum is closely aligned with the 
temporal axis, allowing an anisotropic discretisation to represent accurately 
the Dirac operator on the quark field.

Implementing an anisotropic programme incurs a number of computational overheads not associated with the 
isotropic lattice. The ratio of scales, $\xi=a_s/a_t$ determined by studying a physical
long-distance probe depends on bare parameters in the lattice action. While this
dependence is straightforward to establish at the tree-level of perturbation theory,
quantum corrections can occur at higher orders. In the quenched approximation on a 3+1 lattice this is not a
serious additional cost as the tuning can be done {\it post-hoc}. 
More worringly, in Ref.~\cite{Harada:2001ei} it was pointed out that the choice of $\xi$
and its relation to the Wilson parameter, $r$ on anisotropic versions of the
Sheikholeslami-Wohlert (SW) action
could introduce ${\cal O}(a_sm_q)$ errors. It is exactly errors of this type that the 
anisotropic action seeks to avoid and the appearance of these terms represents a serious tuning problem. 

In this paper we use an action, specifically designed for highly anisotropic lattices
ie. $\xi\geq5$. By applying different improvement terms in the spatial and temporal
directions the action is both doubler and ${\cal O}(a_sm_q)$ error free. This opens up the
possibility of simulating directly at the bottom quark mass using a relativistic action. 
In addition, in this feasibility study the speed of light was determined at $\approx$ 1\%
accuracy. This precision was governed by finite statistics and could certainly be improved upon. 

The paper is organised as follows. The construction of the action is described in 
Section~\ref{sec:quark-action}. Section~\ref{sec:heavyquarks-and-Aria} compares this
action with the sD34 action proposed in Ref.~\cite{Hashimoto:2003fs} and details some
analytic results. Results from a study of the dispersion relations and the mass-dependence 
of the speed of light are described in Section~\ref{sec:results}. Some preliminary
results of this study have appeared in Ref.~\cite{Foley:2003dh}. Our conclusions and a
discussion of future work are contained in Section~\ref{sec:conclusions}. 

%%%%%%%%%%%%%%%%%%%%%%%%%%%%%%%%%%%%%%%%%%%%%%%%%
\section{Designing highly anisotropic actions}
\label{sec:quark-action}
%%%%%%%%%%%%%%%%%%%%%%%%%%%%%%%%%%%%%%%%%%%%%%%%%
We begin by considering a Wilson-type action with Symanzik improvement to remove
discretisation errors. Full ${\cal O}(a)$-improvement requires a clover term and a field
rotation, given by 
\begin{eqnarray}
  \psi       &=& [1-\frac{ra}{4}(\Dslash -m)]\psi^\prime ,
	\label{eqn:Psi_iso_rotation}\\
  \bar{\psi} &=& \bar{\psi^\prime}[1-\frac{ra}{4}(\Dslash -m)] ,
	\label{eqn:Psibar_iso_rotation}
\end{eqnarray}
where $a$ is the lattice spacing on an isotropic lattice and $r$ is the usual Wilson parameter. 
The rotation described by Eqs.~(\ref{eqn:Psi_iso_rotation}) and (\ref{eqn:Psibar_iso_rotation})
preserves locality and maintains a positive transfer matrix so that ghost
states do not arise in a calculation of the free fermion propagator. 
However, in an anisotropic implementation of this action, when $a_t$ is made very small,
these rotations may lead to the reappearance of doublers, an undesirable side-effect of the anisotropy. 

We would like to maintain the useful properties of actions with nearest neighbour temporal
interactions only. In particular, the positivity of the transfer matrix guarantees that 
effective masses approach a plateau from above. 
Therefore, to construct an action suitable for high anisotropies we begin by
applying field rotations in the temporal direction only, 
rewriting Eqs.~(\ref{eqn:Psi_iso_rotation}) and (\ref{eqn:Psibar_iso_rotation}) as 
\begin{eqnarray}
  \psi       &=& [1-\frac{ra_t}{4}(\gamma_0D_0 -m)]\psi^\prime ,
	\label{eqn:aniso_psi_rotation}\\
  \bar{\psi} &=& \bar{\psi^\prime}[1-\frac{ra_t}{4}(\gamma_0D_0 -m)] .
  \label{eqn:aniso_psibar_rotation}
\end{eqnarray}
This leads to a new action in which the temporal and spatial directions are treated
differently. Having applied the rotations of Eqn. (\ref{eqn:Psibar_iso_rotation})
the continuum action is given by
\begin{equation}
S^\prime = \bar{\psi^\prime}M_r\psi^\prime -\frac{ra_t}{2}\bar{\psi^\prime}
           \left( D_0^2 - \frac{g}{2}\epsilon_iE_i\right)\psi^\prime .
	\label{eqn:cont_action}
\end{equation}
where $M_r=\mu_r\gamma_iD_i + \gamma_0D_0 + \mu_rm$ and $\mu_r=(1+\frac{1}{2}ra_tm)$. 
At the tree-level, the rotations described in Eqs.~(\ref{eqn:aniso_psi_rotation}) and 
(\ref{eqn:aniso_psibar_rotation}) do not generate a spatial clover term. As a result the
($\mathbf{\sigma\cdot B}$) term does not appear in Eq.~(\ref{eqn:cont_action}). The
chromoelectric field, $E_i$ is defined as 
\begin{equation}
igE_i = [D_i,D_0] ,
\end{equation}
and $\epsilon_i\equiv\sigma_{i0}$ is given by $\epsilon_i=\displaystyle{\frac{1}{2i}}[\gamma_i,\gamma_0]$.

The temporal doublers are removed by discretising the $D_0^2$ term in the usual way. However, with 
no spatial rotation the spatial doublers remain and must be treated separately. 
They are removed by adding a higher-order, irrelevant operator to the action. This 
was first suggested by Hamber and Wu in Ref.~\cite{Hamber:1983qa}.
The simplest such operator is a spatial $D^4$ term which is added {\it ad hoc} to the Dirac operator 
giving an action,
\begin{equation}
S^\prime = \bar{\psi^\prime}M_r\psi^\prime -\frac{ra_t}{2}\bar{\psi^\prime}
           \left( D_0^2 - \frac{g}{2}\epsilon_iE_i\right)\psi^\prime 
            + sa_s^3\bar{\psi^\prime}\sum_iD^4_i\psi^\prime .
	\label{eqn:continuum-action}
\end{equation}
This approach has previously been discussed in detail in Ref.~\cite{Peardon:2002sd}. 
In this formulation, $s$ is a Wilson-like parameter which is chosen such that the doublers receive 
a sufficiently large mass. The discretisation of the action in Eq. (\ref{eqn:continuum-action}) is now 
straightforward. Only the $\gamma_iD_i$ term requires an improved discretisation since the 
simplest discretisation would lead to ${\cal O}(a_s^2)$ errors. For this case we write
\begin{eqnarray}
\Delta_{imp}^{(1)}\phi (x) &=& \frac{1}{a}\left\{ \frac{2}{3}\left[ \phi (x+a)-\phi (x-a)\right]\right.\nonumber\\
                           & & \left. -\frac{1}{12}\left[\phi (x+2a)-\phi (x-2a)\right] \right\}
	\label{eqn:improv-deriv}
\end{eqnarray} 
and similarly the (unimproved) discretisations of $\partial$, $\partial^2$ and $\partial^4$ are
\begin{eqnarray}
\Delta^{(1)}\phi (x) &=& \frac{1}{2a}\left\{ \phi (x+a)-\phi (x-a)\right\} ,\\
\Delta^{(2)}\phi (x) &=& \frac{1}{a^2}\left\{ \phi (x+a)+\phi (x-a)-2\phi\right\} ,\\
\Delta^{(4)}\phi (x) &=& \frac{1}{a^4}\left\{\left[\phi (x+2a)+\phi (x-2a)\right]\right.\nonumber\\ 
                     &-&  \left. 4\left[\phi (x+a)-\phi (x-a)\right]+6\phi(x)\right\}.
\end{eqnarray}
The corresponding gauge covariant derivatives, $D$, $D^2$ and $D^4$ respectively are constructed by including 
link variables in the usual way. The chromoelectric field is discretised by a clover term with plaquettes 
in the three space-time planes only
\begin{equation}
gE_i = \frac{1}{\xi a_t^2}\frac{1}{u_s^2u_t^2}\frac{1}{8i} 
          \left\{ \Omega_i(x) - \Omega^\dagger_i(x)\right\} ,
\end{equation}
with 
\begin{eqnarray}
\Omega_i(x) &=&  U_i(x)U_t(x+\hat{\imath})
	 U_i^\dagger (x+\hat{t})U_t^\dagger (x) \nonumber \\
     &+&  U_t(x)U_i^\dagger (x-\hat{\imath}+\hat{t})U_t^\dagger (x-\hat{\imath})
         U_i(x-\hat{\imath}) \nonumber \\
     &+&  U_i^\dagger (x-\hat{\imath})U_t^\dagger (x-\hat{\imath}-\hat{t})
	 U_i(x-\hat{\imath}-\hat{t})U_t(x-\hat{t})  \nonumber \\
     &+&  U_t^\dagger (x-\hat{t})U_i(x-\hat{t})U_t(x+\hat{\imath}-\hat{t})U_i^\dagger(x)  .\nonumber\\
\end{eqnarray}
$u_s$ and $u_t$ are the mean-link improvement parameters. $u_s$ is determined from the spatial plaquette 
and $u_t$ is set to unity. 
At the accuracy of the action constructed here no improvement is required. Finally, including the gauge 
fields and the mean-link improvement factors the lattice fermion matrix is given by,
\begin{widetext}
\begin{eqnarray}
M_{\rm ARIA}\psi (x) &=& \frac{1}{a_t}\left\{\left(\mu_rma_t+\frac{18s}{\xi}+r+\frac{ra_t^2g}{4}\epsilon_iE_i\right)
		\psi (x) - \frac{1}{2u_t}\left[ (r-\gamma_0)U_t(x)\psi (x+\hat{t}) 
	         +(r+\gamma_0)U_t^\dagger(x-\hat{t})\psi (x-\hat{t})\right] \right.\nonumber \\
	    & & \left. -\frac{1}{\xi_q}\sum_i\left[ \frac{1}{u_s}(4s-\frac{2}{3}\mu_r\gamma_i)U_i(x)
        	\psi (x+\hat{\imath}) +\frac{1}{u_s}(4s+\frac{2}{3}\mu_r\gamma_i)U_i^\dagger(x-\hat{\imath})
		\psi (x-\hat{\imath}) \right.\right. \nonumber \\ 
            & & \left.\left. -\frac{1}{u_s^2}(s-\frac{1}{12}\mu_r\gamma_i)U_i(x)U_i(x+\hat{\imath})
	        \psi (x+2\hat{\imath}) - \frac{1}{u_s^2}
		(s+\frac{1}{12}\mu_r\gamma_i)U_i^\dagger (x-\hat{\imath})U_i^\dagger
	        (x-2\hat{\imath}) \psi (x-2\hat{\imath})\right] \right\} .
	\label{eqn:ARIA-M_L}
\end{eqnarray}
\end{widetext}
At the tree-level, the fermion anisotropy $\xi_q$ is given by the ratio of scales, $\xi= a_s/a_t$. We call the action
described here ARIA for Anisotropic, Rotated, Improved Action. It is classically 
improved to ${\cal O}(a_t, a_s^3)$. 
%%%%%%%%%%%%%%%%%%
\section{Heavy quarks with ARIA}
\label{sec:heavyquarks-and-Aria}
%%%%%%%%%%%%%%%%%%
The precision calculation of the glueball spectrum on coarse
lattices~\cite{Morningstar:1999rf} suggests that heavy hadronic quantities
would also benefit from the anisotropic formulation. The correlation functions for heavy-heavy and 
heavy-light mesons fall off rapidly with time and it can be difficult to isolate a convincing plateau over a 
reasonable number of timeslices. A lattice with fine temporal direction in principle solves this problem by 
providing a large number of timeslices over which the time-dependence can be resolved. Improved Wilson actions on 
anisotropic lattices have been used to study a range of heavy flavour
physics including charmonium and bottomonium 
spectroscopy~\cite{Okamoto:2001jb,Liao:2001yh,Chen:2000ej}, heavy-light and hybrid 
spectra~\cite{Edwards:2003cd,Matsufuru:2002vh,Luo:2002rz,Harada:2002ii} and also
heavy-light semileptonic decays~\cite{Shigemitsu:2002wh}. 

In these calculations, currents are improved using rotations, which are applied identically in all four space-time 
directions and the Wilson parameter in the spatial direction is usually chosen to be 
either $r_s=1/\xi$~\cite{deForcrand:1999df,Umeda:2000ym} or 
$r_s=1=r_t$~\cite{Klassen:1998jf,Klassen:1998fh,Chen:2000qj,AliKhan:2000bv}. 
However, it was pointed out in Ref.~\cite{Harada:2001ei} that simulations with anisotropic Wilson-type 
actions may include ${\cal O}(a_sm_q)$ effects. 
Naively, errors of this form are unexpected but they arise in products of  
the Wilson and mass terms in the action. 
In particular, the authors showed that the presence of these artefacts, which appear in radiative corrections, 
depends on the spatial Wilson parameter, $r_s$. The ${\cal O}(a_sm_q)$-dependence 
potentially spoils the benefits of working on an anisotropic lattice, especially at large
quark masses. 

In Ref.~\cite{Hashimoto:2003fs} a different approach was adopted. Since the unwanted ${\cal O}(a_sm_Q)$-dependent 
terms arise from the spatial Wilson term the authors propose an anisotropic D234-type 
action~\cite{Alford:1997nx} may be more suitable. In this case a rotation term is 
applied in the temporal direction only, removing the temporal doublers. 
Spatial doublers are removed by adding an irrelevant, dimension-four term to the Dirac operator. 
The authors showed to one-loop order in perturbation theory, that this so-called ``sD34'' action 
does not suffer from ${\cal O}(a_sm_q)$ terms.
Comparing the ARIA action proposed in Section~\ref{sec:quark-action} and the sD34 action from 
Ref.~\cite{Hashimoto:2003fs} we see that these are the same, up to ${\cal O}(a_t)$ improvement. 

The D234 quark action on an anisotropic lattice~\cite{Alford:1997nx} is be written
\begin{equation}
S_{D234} = a_ta_s^3\sum_{x}\bar{\psi}(x) M\psi(x) ,
\end{equation}
and writing $M$ in the notation of Ref.~\cite{Hashimoto:2003fs}
\begin{eqnarray}
	M &=& m_0 + \sum_\mu\nu_\mu\gamma_\mu\nabla_\mu(1-b_\mu a_\mu^2\Delta_\mu) 
	       - \frac{1}{2}a_t\left(\sum_\mu r\Delta_\mu   \right.\nonumber\\
	  & & \left. + \sum_{\mu <\nu}c_{\rm SW}^\mu\sigma_{\mu\nu}F_{\mu\nu}\right) 
		+\sum_\mu\nu_\mu d_\mu a_\mu^2\Delta_\mu^2 .
	\label{eqn:SD234}
\end{eqnarray}
The sD34 action is a special case of this action in which the parameters have the following values
\begin{equation}
	\begin{array}{lll}
(\nu_0,\nu_i) = (1,\nu); & (b_0,b_i) =(0,\frac{1}{6}); & \hspace{-1cm}(d_0,d_i) = (0,\frac{1}{8});\\
(r_0,r_i) = (r_t,0); & (c_{\rm SW}^0,c_{\rm SW}^i) = (0,c_{\rm SW}). &\\
	\end{array}
\end{equation}	        
and $\nu = (1+\frac{1}{2}r_ta_tm_0)$. Substituting in Eq. (\ref{eqn:SD234}) gives
\begin{eqnarray}
	M_{\rm sD34} &=& m_0 + \sum_i\nu\gamma_i\nabla_i\left( 1-\frac{1}{6}a_s^2\Delta_i\right) 
			     + \gamma_0\nabla_0 \nonumber \\
                     & &\hspace{-1cm} - \frac{a_t}{2}\left( r\Delta_0 + c_{\rm SW}^t\sigma_{i0}F_{i0}\right) 
			 + \frac{1}{8}a_s^3\sum_i\nu\Delta_i^2 ,
	\label{eqn:sD34}
\end{eqnarray}
which is the action we use in our simulations, up to ${\cal O}(a_t)$
improvement. Reexpressing the fermion matrix in our notation,
\begin{eqnarray}
	M_{\rm ARIA} &=& \mu_rm_0 + \sum_i\mu_r\gamma_i\nabla_i\left( 1- \frac{1}{6}a_s^2\Delta_i\right)
			+ \gamma_0\nabla_0 \nonumber\\
	            & & -\frac{a_t}{2}\left( r\Delta_0 -\frac{rg}{2}\sigma_{i0}F_{i0}\right) 
                        + sa_s^3\sum_i\Delta_i^2 ,
	\label{eqn:M-ARIA}
\end{eqnarray}
where $s=1/8$ and $\mu_r=\left( 1+\frac{1}{2}r_ta_tm_0\right)$. 
%%%%%%%%%%%%%%%%%%%%%%%%%%%%%%%%%%%%%%%
\subsection{Analytic results for ARIA}
%%%%%%%%%%%%%%%%%%
In this section the energy-momentum behaviour of the ARIA action is calculated. We begin by presenting results 
for general $r$ and $s$. The free-quark dispersion relation is obtained by solving 
$\det\widetilde{M}_{\rm ARIA} = 0$ in momentum space where $\widetilde{M}_{\rm ARIA}$ is the Fourier
transform of $M_{\rm ARIA}$ in Eq. (\ref{eqn:ARIA-M_L}).
The energy-momentum relation is 
\begin{eqnarray}
	\cosh(E a_{t}) &=& \frac{ r^2 + r\omega(p)}{r^2 - 1}\pm \nonumber\\
                       & & \hspace{-1cm}\frac{\sqrt{(r+\omega (p))^2+(1-r^2)(1+a^2_t\tilde{\mathbf{p}}^2) }}{r^2 - 1}, 
	\label{eqn:cosh_reln}
\end{eqnarray}
where $\omega(p)$ and $\tilde{\mathbf{p}}$ are defined as
\begin{eqnarray}
	\omega(p)          &=& a_{t} \mu_r m_0 + a_t s\sum_i a^3_s \hat{p}^4_i ,\\
	\tilde{\mathbf{p}} &=& \mu_r\bar{p}_i( 1 + \frac{1}{6}a^2_s\hat{p}^2_i ),
\end{eqnarray}
with $\bar{p}_i = \displaystyle{\frac{1}{a_s}}\sin(a_sp_i)$ and $\hat{p}_i=\displaystyle{\frac{2}{a_s}}\sin(a_sp_i/2)$.
Expanding the physical solution in powers of spatial momentum yields  
\begin{equation}
	E^{2}(\mathbf{p}) = M_{1}^{2} + \frac{M_{1}}{M_{2}} \mathbf{p}^2 +
	O(\mathbf{p}^{4}),
	\label{eqn:cont-disp}
\end{equation}  
where $M_1$ is the rest mass, given by 
\begin{widetext}
\begin{equation}
	M_{1} = \frac{1}{a_t}\cosh^{-1}\left(\frac{r^2+\mu_rm_0ra_{t} - 
		\sqrt{1+2\mu_r m_0ra_t+ m_0^2\mu^2_r a^2_t} }{r^2 -1} \right).
		\label{eqn:ARIA-restmass}
\end{equation}
\end{widetext}
The kinetic mass, $M_{2}$ is given by
\begin{widetext}
\begin{equation}
	 \frac{1}{M_2} = \frac{ \mu^2_r a_t } { \sqrt{1 + 2 \mu_r m_0ra_t + m_0^2\mu^2_r a^2_t }} 
			 \left[\left(\frac{r^2+\mu_r m_0ra_t - \sqrt{1+2\mu_r m_0ra_t+m_0^2\mu^2_r
				a^2_t }}{r^2 -1}\right)^2 - 1 \right]^{-\frac{1}{2} }.
	\label{eqn:ARIA-kineticmass}
\end{equation}
\end{widetext}
Eqs.~(\ref{eqn:ARIA-restmass}) and (\ref{eqn:ARIA-kineticmass}) indicate that at the tree-level 
$M_1$ and $M_2$ do not depend on ${\cal O}(a_sm_q)$ terms or on the ratio of scales, $\xi$.

To compare these expressions with the results of other studies,
the particular choice $r=1$ was considered. In this case the lattice ghost (the unphysical solution of 
Eq.~(\ref{eqn:cosh_reln})) disappears and the dispersion relation is given by
\begin{equation}
	4\sinh^{2}\left( \frac{E a_{t}} {2}\right) = \frac{a^{2}_{t}\tilde{\mathbf{p}}^{2}+
					              \omega^{2}(p)} { 1 + \omega(p)},
	\label{eqn:sinh_reln}
\end{equation}
with
\begin{eqnarray}
	M_{1}                &=& \frac{1} {a_{t}} \log(1 + \mu_rm_0a_{t}), \\
	\frac{1} { 2 M_{2} } &=& \frac{\mu_r} {m_0( 2 + \mu_rm_0a_{t}) }.
\end{eqnarray}
where now $\mu_r=(1+\frac{1}{2}a_tm_0)$. 
These expressions are consistent with those obtained in Ref.~\cite{Hashimoto:2003fs} for the sD34 action and in 
Ref.~\cite{El-Khadra:1997mp} for the Fermilab action on an isotropic lattice. 

\begin{figure}[h]
	\centering
 	\includegraphics[width=7cm]{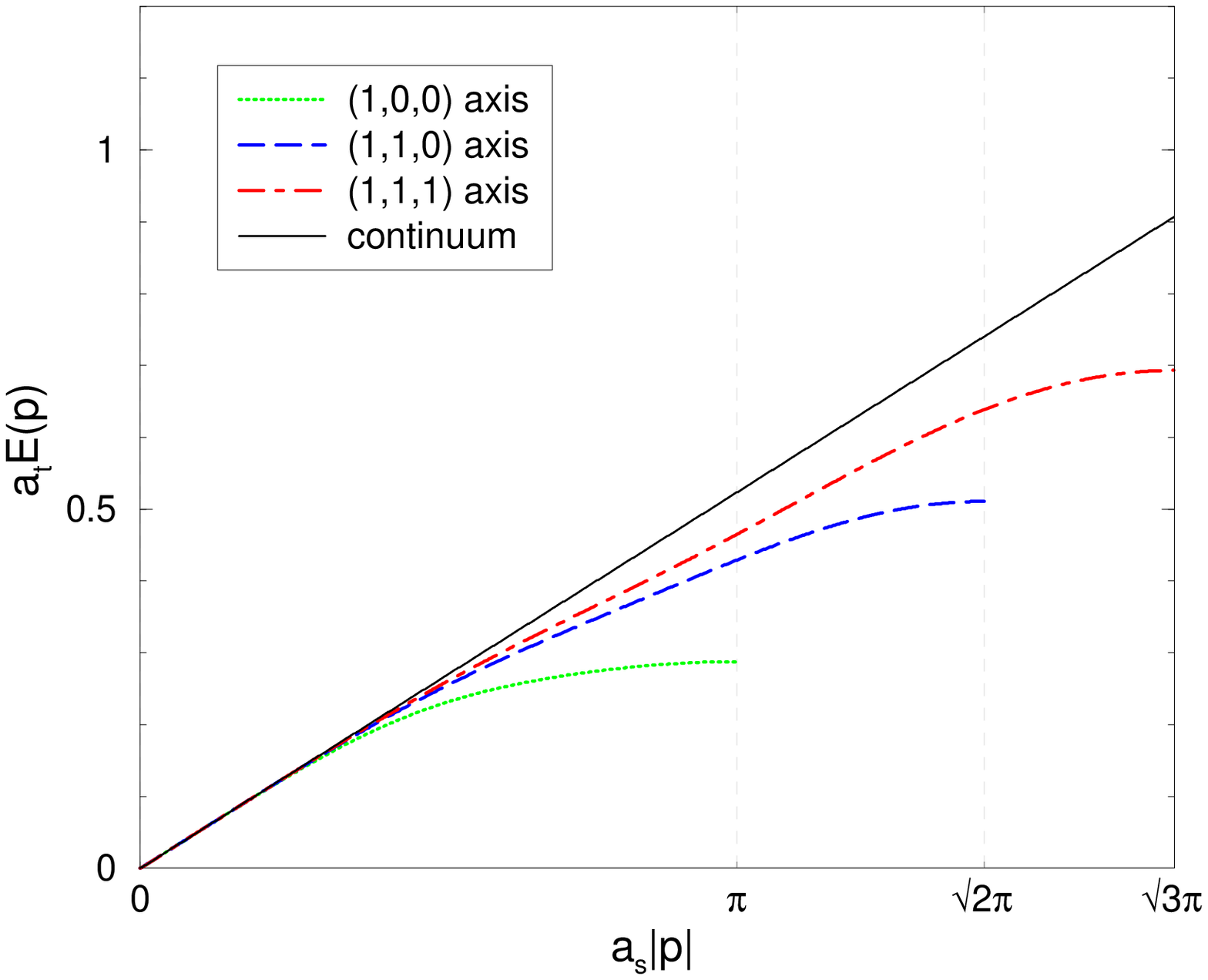}
	\includegraphics[width=7cm]{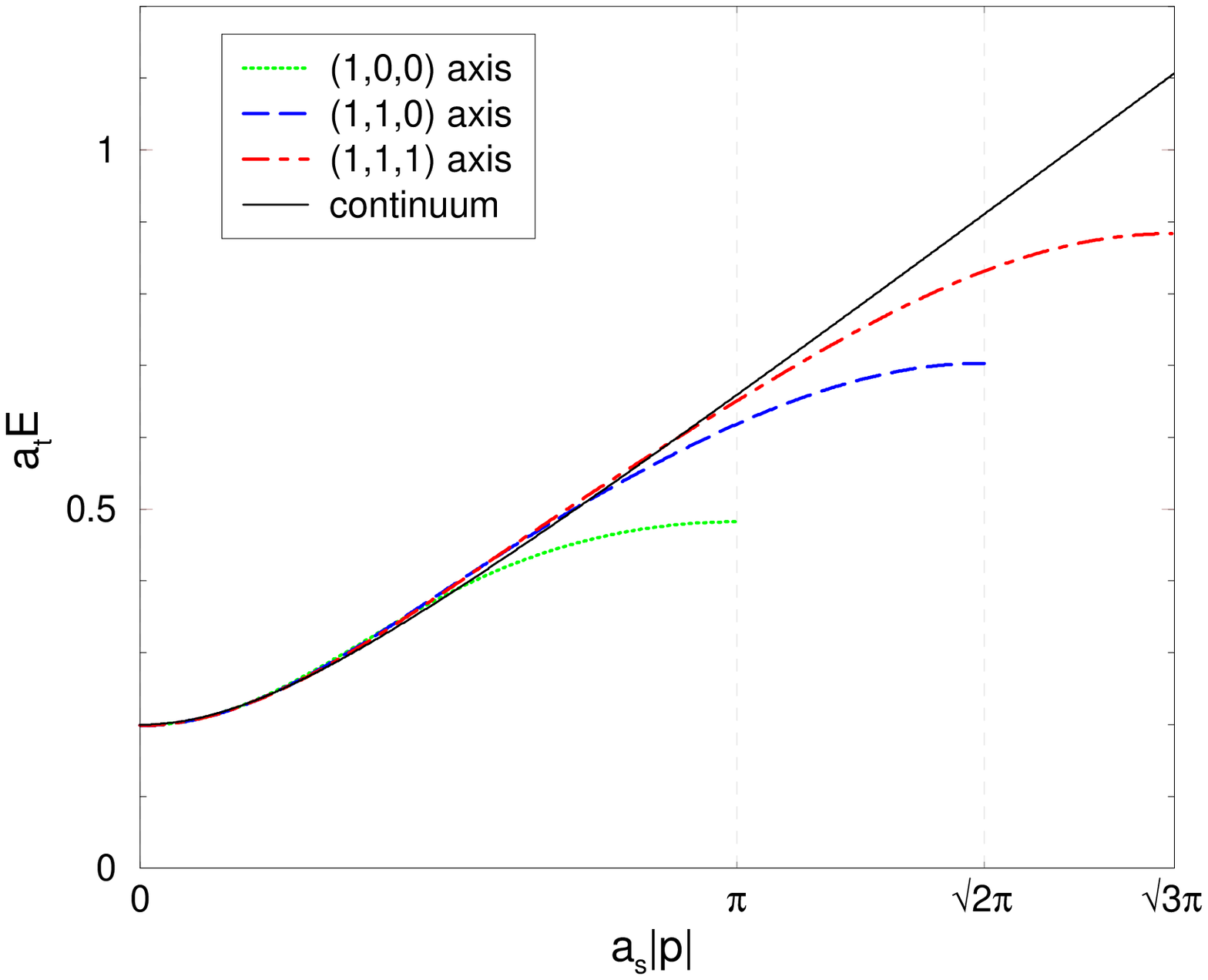}
	\caption{The dispersion relations given by Eq.~\ref{eqn:cosh_reln} with $\xi=6$, $r=1$ and $s=1/8$.
		 The top figure is the massless case while the bottom plot shows the massive
	       	 case, with $a_tm_q=0.2$.}
	\label{fig:disp_relation}
\end{figure}

The free-quark dispersion relations for massless and massive quarks are shown in 
Figure~\ref{fig:disp_relation}. The anisotropy parameter, $\xi$ is six for both cases.  
%%%%%%%%%%%%%%%%%%
In analogy to the traditional Wilson $r$-parameter, the parameter $s$ in this action can in principle take 
any positive value. We chose $s=1/8$ by eye, demanding that the energy-momentum relations do not have negative 
slope for $a_s|p|<\pi$. Since $s$ parameterises a term which removes the spatial doublers and is irrelevant in the 
continuum limit precise tuning is not required. 
%%%%%%%%%%%%%%%%%%%
\section{Results}
\label{sec:results}
%%%%%%%%%%%%%%%%%%%
In this exploratory study the temporal rotations have been omitted which leads to an 
${\cal O}(a_t)$ classical discretisation error. However, since $a_t$ is small in these simulations, 
$a_t \sim 0.04$ fm, the effects should be under control at least when $a_tm_q <
1$. Discarding temporal rotations means the action has no clover term and in addition we
have set $\mu_r=1$. It is planned to include correction terms to remove ${\cal O}(a_t)$
errors in future work.

The ratio of scales is changed in a simulation by quantum 
corrections. Therefore the action parameter must be adjusted so that the ratio of scales
measured from a physical quantity is correct. In a quenched simulation the parameters
$\xi_g$ and $\xi_q$ in the gauge and quark actions may be independently tuned to the
target anisotropy, using different physical probes. 
This is not the case for unquenched simulations where the anisotropy in the gauge and
quark actions must be tuned simultaneously~\cite{Peardon:2002sd}. 

For this study an ensemble of quenched gauge configurations for which $\xi_g$ had already been tuned 
was used. In this case the tuning criterion was that $\xi=6$ when measured from the static interquark 
potential in different directions on the lattice. 
The parameter $\xi_q$ in the fermion action must now be tuned such that its value determined from the 
energy momentum dispersion relation is six. 
At this point we introduce some terminology which makes clear the difference between $\xi_q$, which is 
a parameter in the action, and the slope of the dispersion relation which is a physical observable -- 
usually called the speed of light, $c$. The target anisotropy is six. $\xi_q$ is tuned so
that the speed of light (determined from the slope of the dispersion relation) is unity.

The anisotropic action offers the possibility of precision studies of a 
range of phenomenologically interesting heavy quark quantities in the $D$, $B$, $J/\psi$ and $\Upsilon$
sectors. For this reason it is important to understand the dependence of $\xi_q$ on the heavy quark 
mass used in simulations. In particular, 
a contribution of ${\cal O}(a_sm_q)$ to the renormalised anisotropy would spoil this tuning for charm and 
bottom quark masses. The main result in this section is a study of the mass-dependence of
the speed of light at fixed anisotropy. 
%%%%
\subsection{Simulation parameters}
\label{subsec:sim-params}
The gauge action used in this simulation is a two-plaquette improved action designed for precision 
glueball simulations on anisotropic lattices. A description is given in 
Ref.~\cite{Morningstar:1999dh}. The construction of the fermion action is described in detail 
in Section~\ref{sec:quark-action}. 
Details of the simulation and parameter values are summarised in Table~\ref{tab:latt-details}. 
%%%%%%%%%%%%
\begin{table}[ht]
\begin{center}
\begin{tabular}{c|c}
\hline
\hline
\# gauge configurations & 100 \\
Volume & $10^3\times 120$ \\
$a_s$              & 0.21fm                    \\
$a_s/r_0$	   & 0.4332(11) \\
$\xi=a_s/a_t$      & 6 \\
$a_tm_q$           & -0.04,0.1,0.2,0.3,0.4,0.5,1.0,1.5 \\
\hline
\end{tabular}
\caption{Details of the simulation.}
\label{tab:latt-details}
\end{center}
\end{table}
A broad range of quark masses was investigated, from $a_tm_q =-0.04$ which is close to the 
strange quark on these lattices to heavy quarks with $a_tm_q=1.0$ and $1.5$. Both
degenerate and nondegenerate combinations are considered. The nondegenerate combination is
made with the lightest quark and each of the heavier quarks. Note that
$a_tm_q =-0.04$ corresponds to a positive quark mass since Wilson-type actions 
have an additive mass renormalisation. 
We accumulated data at spatial momenta (0,0,0), (1,0,0), (1,1,0) and (1,1,1), in units of 
$2\pi /a_sL$, averaging over equivalent momenta. 
%%%

It is worth noting that all the gauge configurations and quark propagators used in this
study were generated on Pentium IV workstations. Generating the lightest quark propagators (close to
the strange quark mass) required approximately one week on a single processor. At this
quark mass no exceptional configurations were seen. 
%%%%%%%%%%%%
\subsection{Effective masses}
\label{subsec:effmass}
The success of anisotropic lattice methods is predominantly due to the increased resolution in the 
temporal direction. The fineness of the lattice in this direction is particularly useful when 
determining heavy mass quantities whose signal to noise ratio decreases rapidly. 
The increase in resolution also leads to reduced statistical errors in effective masses since 
fits can be made to longer time ranges than is usually possible with an isotropic
lattice. For the same reason, the fitted values tend to be less sensitive to fluctuations
of one or two points in the chosen fit range. 

\begin{figure}[ht]
	\centering
 	\includegraphics[width=7cm]{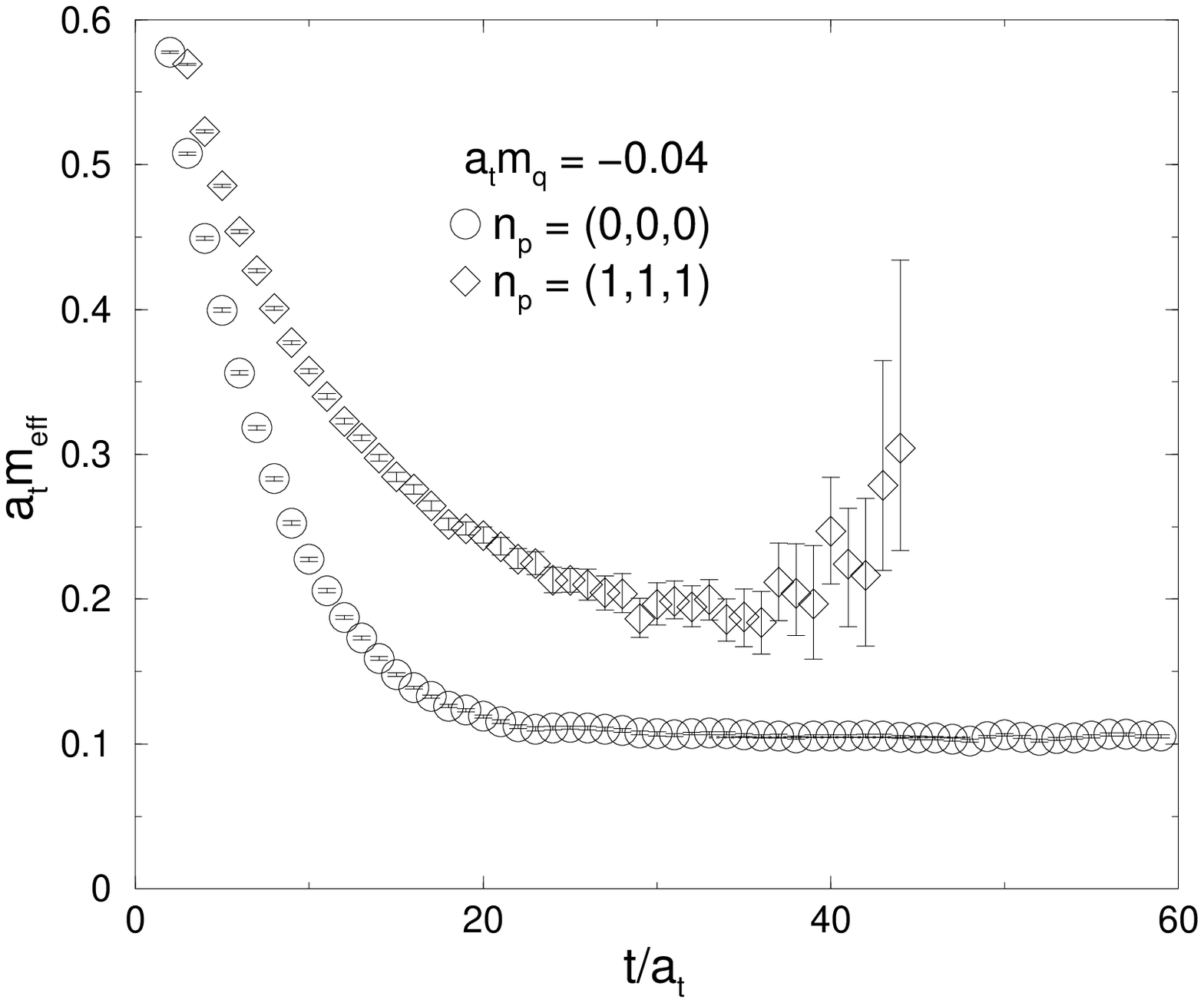}
 	\includegraphics[width=7cm]{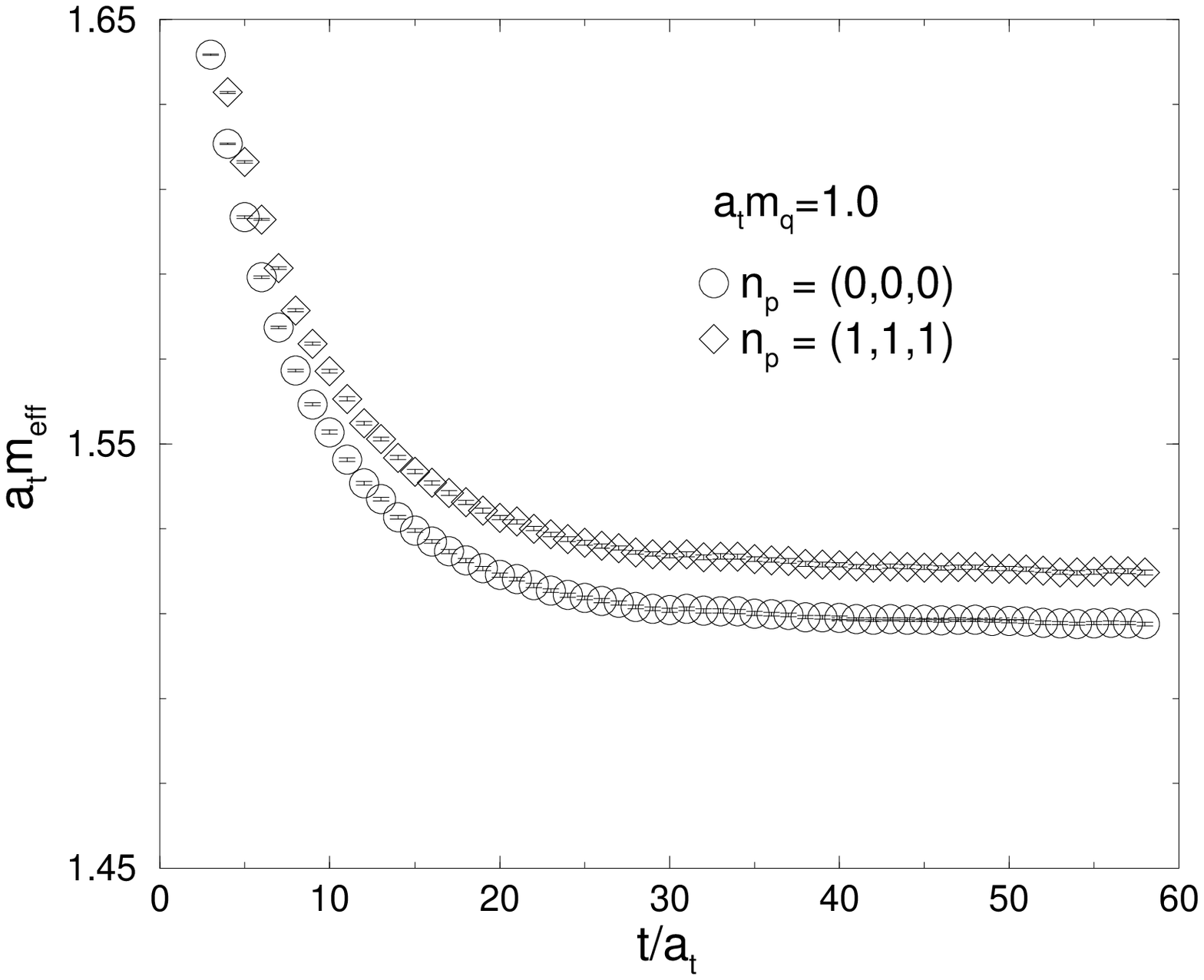}
	\caption[fig:PSeff-mass]{Pseudoscalar meson effective mass plots. The two plots
 	indicate that very good fits can be made for a wide range of quark masses and momenta. 
	The top plot shows the effective mass of the lightest meson made from a 
	degenerate combination of quarks with $a_tm_q=-0.04$ for zero momentum and for
 	three units of momentum in lattice units. The second plot is the analogous case
 	for $a_tm_q=1.0$.}
	\label{fig:PSeff-mass}
\end{figure}

\begin{figure}[ht]
	\centering
 	\includegraphics[width=7cm]{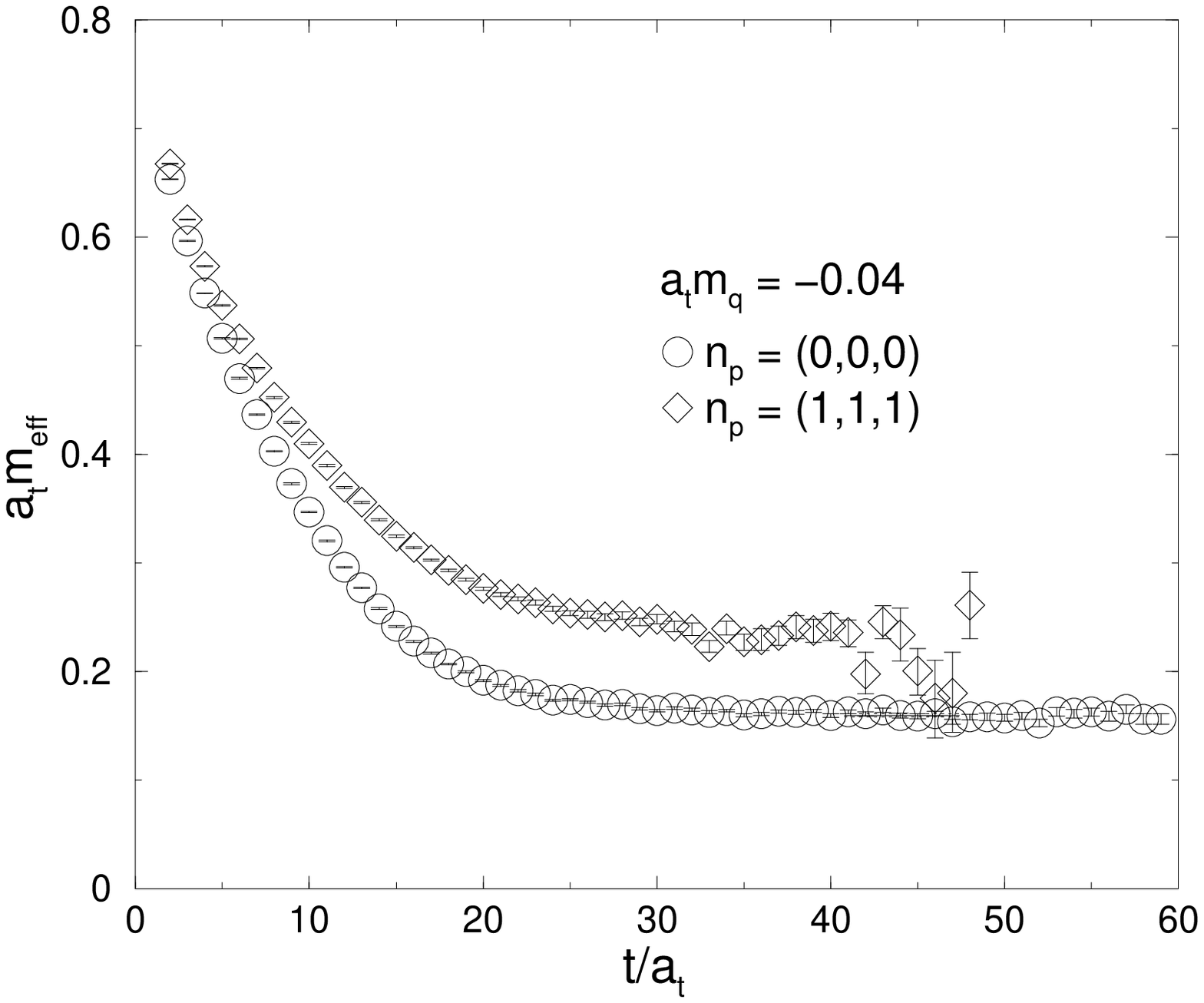}
 	\includegraphics[width=7cm]{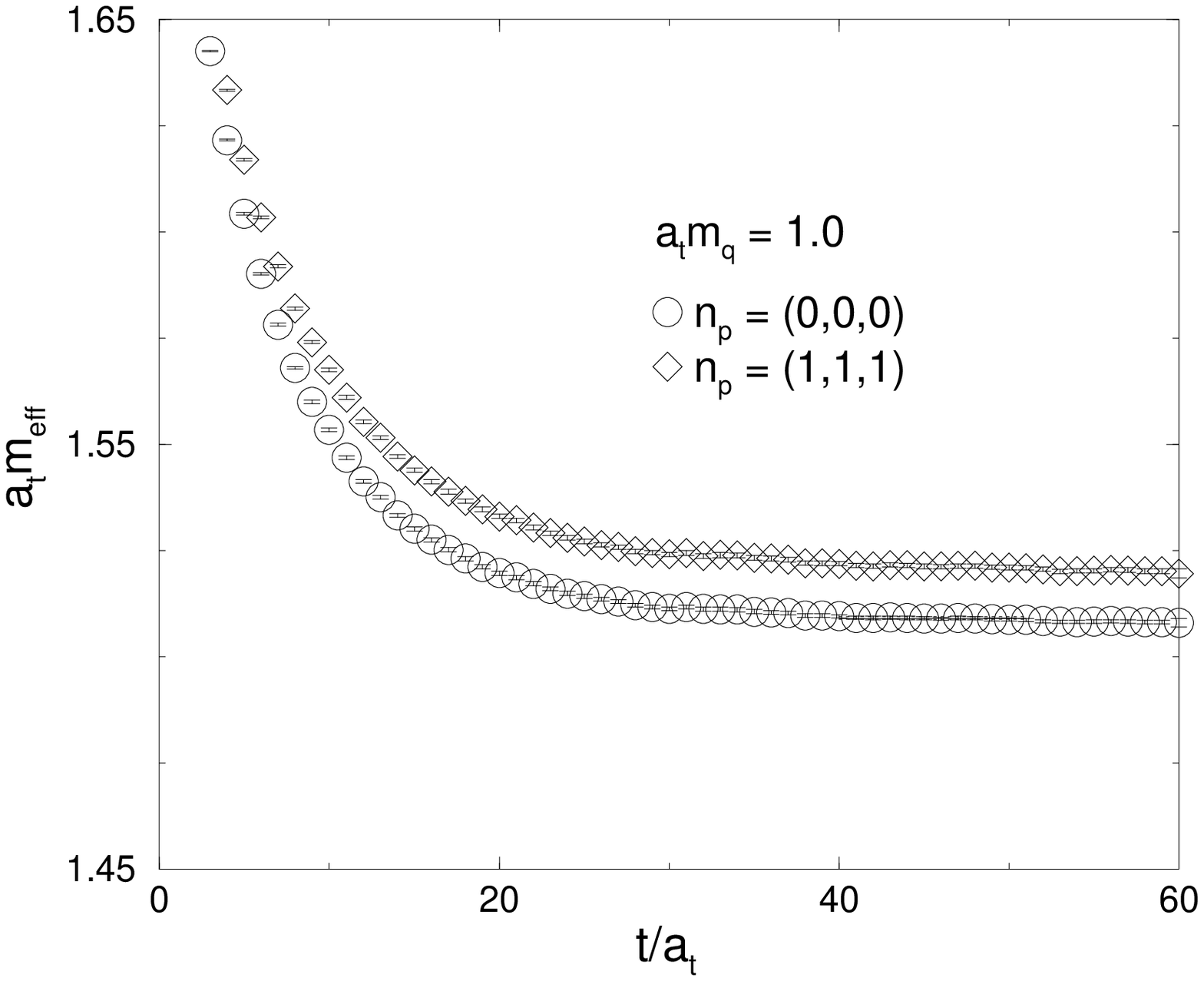}
	\caption[fig:Veff-mass]{Vector meson effective mass plots. As in the pseudoscalar
 	case shown in Figure~\ref{fig:PSeff-mass}, good fits are achieved over a large number
 	of timeslices for all the quark masses considered in this study. The top plot
 	shows the lightest degenerate vector while the bottom plot shows the same result
 	for $a_tm_q=1.0$.}
	\label{fig:Veff-mass}
\end{figure}

In this study the effective masses were determined using single cosh fits with a $\chi^2$ minimization algorithm. The 
signal to noise ratio was enhanced by using four sources, distributed across the lattice
at timeslices 0, 30, 60 and 90. The average of these results was used in the effective mass fits. The 
statistical errors shown are calculated from 1000 bootstrap samples in each fit. 
Figure~\ref{fig:PSeff-mass} shows four effective mass plots. The first plot is the
pseudoscalar meson with degenerate quarks at the lightest mass for zero momentum and for
momentum of $(1,1,1)$ in lattice units, $2\pi /a_sL$. The second plot is the analogous case for the
degenerate combination of quarks with $a_tm_q=1.0$. In all cases a clear plateau, over a large number of timeslices is 
observed. The fits to effective masses of the non-degenerate mesons are equally good and
in all cases the fit range is ten or more timeslices with a $\chi^2$ per degree of freedom
($\chi^2/N_{df}$) $\sim 1$.
%%%%%%%%%%%%%%%%%
%%%%%%%%%%%%%%%%%
In Figure~\ref{fig:Veff-mass} the equivalent results for vector mesons are presented. Once
again, the lightest and heaviest degenerate combinations of quark masses considered are shown and very
good fits are possible in both cases. 
%%%%%%%%%%%%%%%%%
%%%%%%%%%%%%%%%%%%%%%%%%%%%%%%%%%%%%%%%%
\subsection{The renormalised anisotropy}
\label{subsec:renorm-aniso}
The slope of the energy momentum dispersion relation is determined nonperturbatively 
and compared to the target anisotropy, $\xi=6$. We are 
interested in both the precision of the determination and the deviation of the speed of light from unity.
The wide range of quark masses used in this simulation ($a_tm_q=-0.04$ to $a_tm_q=1.5$) allow us to study 
the mass-dependence of the renormalisation. We also examine the difference between the anisotropy
determined from particles with degenerate and non-degenerate quark content. 

\begin{figure}[ht]
	\centering
	\includegraphics[width=7cm]{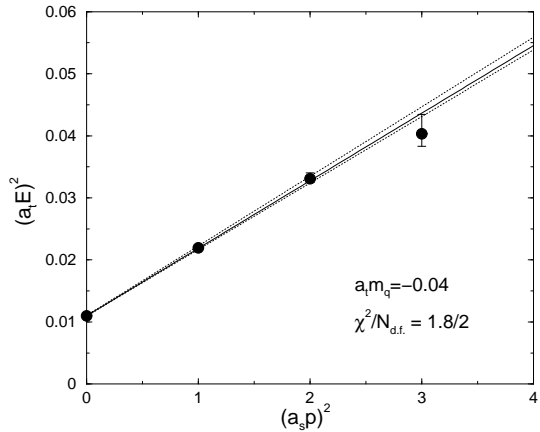}
	\caption[fig:lightquark-dispreln]{The energy-momentum for the lightest degenerate meson in this 
					  simulation. The bare quark mass is $a_tm_q=-0.04$.}
	\label{fig:lightquark-disp}
\end{figure}

\begin{figure}[ht]
	\centering
	\includegraphics[width=7cm]{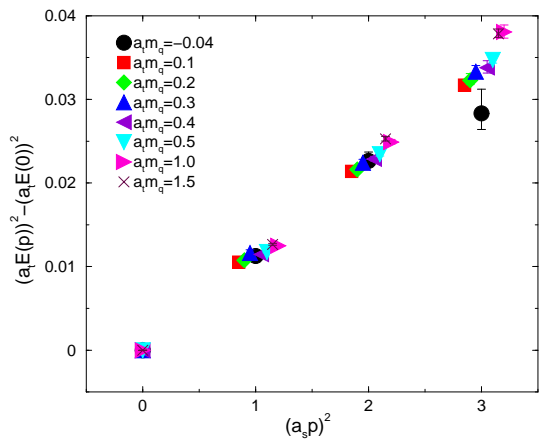}
	\caption[fig:disp-reln]{The energy momentum dispersion relation for the
	pseudoscalar meson at all the masses simulated. The lightest point, $a_tm_q=-0.04$
	is close to the strange quark while the heaviest mass is close to the bottom
	mass. The points have been shifted about their momentum value to make the plot
	easier to read.}
	\label{fig:disp-reln}
\end{figure}

To begin, the dispersion relation was determined for a pseudoscalar meson made from the
lightest quarks in this simulation, $a_tm_q=-0.04$ and with an input anisotropy,
$\xi_q=6.0$. The value of $c$ determined from the dispersion relation was used to
determine the tuned value of the anisotropy, $\xi_q=6.17$ and the calculation 
repeated. The resulting dispersion relation is shown in Figure~\ref{fig:lightquark-disp}. The
subsequent value of $c$ determined from this data is $1.02\pm 0.01$. 
%%%%%%%%%%%%%%%%%
%%%%%%%%%%%%%%%%%
This value of $\xi_q =6.17$ was then used in simulations for a range of quark masses, 
$0.1\leq a_tm_q\leq 1.5$. A representative sample of the energy-momentum 
dispersion relations for this range of quark masses and particles is shown in Figure~\ref{fig:disp-reln}. 
The plot shows very good linear dispersive behaviour. This relativistic dispersion
relation persists for both degenerate and non-degenerate quark combinations in  
pseudoscalar and vector particles at all masses. 

The mass-dependence of the speed of light is given by the difference in the 
slopes for the different masses. Figure~\ref{fig:disp-reln} shows that this dependence is
mild. The lightest mass, close to the strange quark mass, is the noisiest and the
statistical errors increase with increasing momentum, as expected. It should be noted that
the quark propagators used in this study are generated with point sources and the use of smearing
techniques is expected to improve the signal for this and
lighter quark masses. In addition the advantages of stout link gauge backgrounds~\cite{Morningstar:2003gk}
will be investigated in further studies. 

\begin{figure}[ht]
	\centering
	\includegraphics[width=7cm]{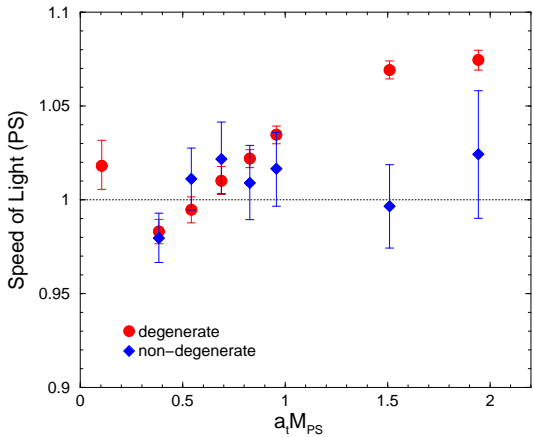}
	\caption[fig:PSrenorm-aniso]{The mass-dependence of the speed of light determined
	from the pseudoscalar (PS) dispersion relations for fixed $\xi_q=6.17$. The plot shows
	mesons with both degenerate and non-degenerate quark mass combinations plotted as a
	function of the degenerate meson mass, in units of the temporal lattice spacing.}
	\label{fig:PSrenorm-aniso}
\end{figure}

\begin{figure}[ht]
	\centering
	\includegraphics[width=7cm]{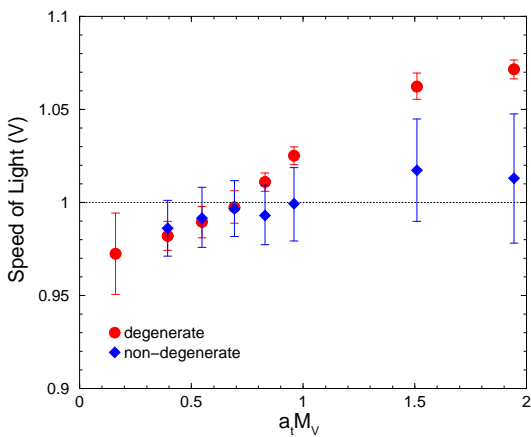}
	\caption[fig:Vrenorm-aniso]{The mass-dependence of the speed of light determined
	from the vector (V) dispersion relations with $\xi_q=6.17$. Both degenerate and non-degenerate mass
	combinations are shown, plotted as a function of the degenerate vector mass as in
	Figure~\ref{fig:PSrenorm-aniso}.}
	\label{fig:Vrenorm-aniso}
\end{figure}

In Tables~\ref{tab:hhdata} and ~\ref{tab:hldata} we show the speed of light determined from
the slope of the dispersion relation for each mass in the simulation. The $\chi^2/N_{d.f.}$
for these fits also is shown. Results for both pseudoscalar and vector 
mesons are given and the ground state masses extracted in the fitting procedure 
described in Section~\ref{subsec:effmass} are listed. 
%%%%%%%%%%%%%%%%%%%%%%%%%%%%%%
\begin{table}[ht]
\begin{center}
\begin{tabular}{c| ccc| ccc}
\hline
 & \multicolumn{3}{c}{Pseudoscalar} & \multicolumn{3}{c}{Vector} \\
\hline
$a_tm_q$ & $a_tM_{PS}$  & $c$ & $\chi^2/N_{df}$ & $a_tM_V$  & $c$ & $\chi^2/N_{df}$ \\ 
\hline
-0.04 & $0.1045^{+5}_{-5}$ & $1.02^{+1}_{-1} $ & $6.3/2$ 
		 & $0.161^{+2}_{-2}$ & $0.97^{+2}_{-2}$ & $0.66/2$\\
0.10  & $0.3831^{+4}_{-4}$ & $0.983^{+6}_{-7}$ & $2.8/2$  
		 & $0.3934^{+4}_{-4}$ & $0.982^{+8}_{-8}$ & $ 2.1/2$\\
0.20  & $0.5418^{+3}_{-4}$ & $0.995^{+7}_{-7}$ & $0.33/2$ 
		 & $0.5472^{+4}_{-4}$ & $0.990^{+8}_{-8}$ & $ 2.1/2$\\
0.30  & $0.6887^{+4}_{-4}$ & $1.010^{+8}_{-7}$ & $2.4/2$  
		 & $0.6924^{+4}_{-4}$ & $0.997^{+9}_{-9}$ & $ 4.5/2$\\
0.40  & $0.8269^{+4}_{-4}$ & $1.022^{+5}_{-5}$ & $0.65/2$ 
		 & $0.8294^{+4}_{-4}$ & $1.011^{+5}_{-5}$ & $ 2.3/2$\\
0.50  & $0.9569^{+4}_{-4}$ & $1.035^{+5}_{-5}$ & $1.3/2$  
		 & $0.9587^{+4}_{-4}$ & $1.025^{+5}_{-5}$ & $ 1.6/2$\\
1.00  & $1.5086^{+3}_{-3}$ & $1.069^{+5}_{-5}$ & $1.3/2$  
		 & $1.5092^{+3}_{-3}$ & $1.072^{+5}_{-5}$ & $ 1.2/2$\\
1.50  &	$1.9428^{+3}_{-3}$ & $1.075^{+5}_{-5}$ & $0.081/2$	
		 & $1.9431^{+3}_{-4}$ & $1.072^{+5}_{-5}$ & $0.058/2$\\	         
\hline
\end{tabular}
	\caption{The ground state pseudoscalar and vector masses with degenerate
	quarks. The speed of light determined from the dispersion relation for each quark
	mass is shown with the associated $\chi^2/N_{d.f.}$. The errors in all cases are
	statistical only. The parameter, $\xi_q$ is fixed in these simulations to $6.17$,
	its value determined from the dispersion relation of the lightest degenerate
	pseudoscalar meson.}
	\label{tab:hhdata}
\end{center}
\end{table}
%%%%%%%%%%%%%%%%%%%%%%%%%%%%%%
\begin{table}[b]
\begin{center}
\begin{tabular}{c| ccc| ccc}
\hline
 & \multicolumn{3}{c}{Pseudoscalar} & \multicolumn{3}{c}{Vector} \\
\hline
$a_tm_q$ & $a_tM_{PS}$  & $c$ & $\chi^2/N_{df}$ & $a_tM_V$  & $c$ & $\chi^2/N_{df}$ \\ 
\hline
0.1 & $0.2610^{+6}_{-6}$ & $0.98^{+1}_{-1}$ & $0.23/2$ & $0.2802^{+8}_{-8}$ & $0.98^{+2}_{-2}$ & $ 0.19/2$\\
0.2 & $0.3466^{+6}_{-6}$ & $1.01^{+2}_{-2}$ & $0.56/2$ & $0.3601^{+8}_{-8}$ & $0.99^{+2}_{-2}$ & $ 0.64/2$\\
0.3 & $0.4254^{+7}_{-7}$ & $1.02^{+2}_{-2}$ & $2/2   $ & $0.4351^{+8}_{-8}$ & $1.00^{+2}_{-2}$ & $ 0.45/2$\\
0.4 & $0.4987^{+7}_{-7}$ & $1.01^{+2}_{-2}$ & $1.5/2 $ & $0.5056^{+8}_{-9}$ & $0.99^{+2}_{-2}$ & $ 1.4/2$\\
0.5 & $0.5668^{+8}_{-8}$ & $1.02^{+2}_{-2}$ & $1.7/2 $ & $0.5720^{+9}_{-9}$ & $1.00^{+2}_{-2}$ & $ 1.6/2$\\
1.0 & $0.8521^{+10}_{-10}$ & $1.00^{+2}_{-2}$ & $2.6/2 $ & $0.854^{+1}_{-1}$ & $1.02^{+3}_{-3}$ & $ 0.62/2$\\
1.5 & $1.074^{+1}_{-1}$ & $1.02^{+3}_{-3}$ & $2.1/2 $ & $1.075^{+1}_{-1}$ & $1.01^{+3}_{-4}$ & $ 1.8/2$\\
\hline
\end{tabular}
	\caption{The ground state masses of non-degenerate combinations of quark
	masses. In each case the quark mass given is combined with the lightest mass in
	our simulations, $a_tm_q=-0.04$. As in Table~\ref{tab:hhdata} the pseudoscalar and
	vector meson states are shown with the speed of light and the associated
	$\chi^2/N_{d.f.}$. Once again all errors are statistical only and $\xi_q=6.17$.}
	\label{tab:hldata}
\end{center}
\end{table}
%%%%%%%%%%%
The dependence of $c$ on the quark mass in the simulation is shown in
Figures~\ref{fig:PSrenorm-aniso} and~\ref{fig:Vrenorm-aniso}. 
The plots show the speed of light as a function of the meson mass in units of 
$a_t$ for both pseudoscalars and vectors.
It is important to remember that the anisotropy was tuned only once at the lightest
pseudoscalar particle. The plots show good agreement between determinations of $c$ from
degenerate and non-degenerate particles up to $a_tm_q\sim 0.5$, corresponding to $a_tM_{PS}=0.957(5)$ in
Figure~\ref{fig:PSrenorm-aniso}. The charm quark mass on this lattice is close to $a_tm_q=0.2$, implying that 
charm physics is both computationally feasible and requires little parameter tuning at an anisotropy 
of six. Figures~\ref{fig:PSrenorm-aniso} and~\ref{fig:Vrenorm-aniso} also show some quark
mass dependence for degenerate mesons with $a_tm_q\geq 0.5$. 
They also indicate that the agreement between the degenerate and non-degenerate meson physics decreases for 
$a_tm_q\geq 0.5$. This is not unexpected since degenerate mesons with two heavy quarks
(charmonium and bottomonium) have a small Bohr radius and can suffer large
discretistation effects. The non-degenerate mesons ($D$ and $B$ mesons) do not have such a
problem. 

\begin{figure}[ht]
	\centering
	\includegraphics[width=7cm]{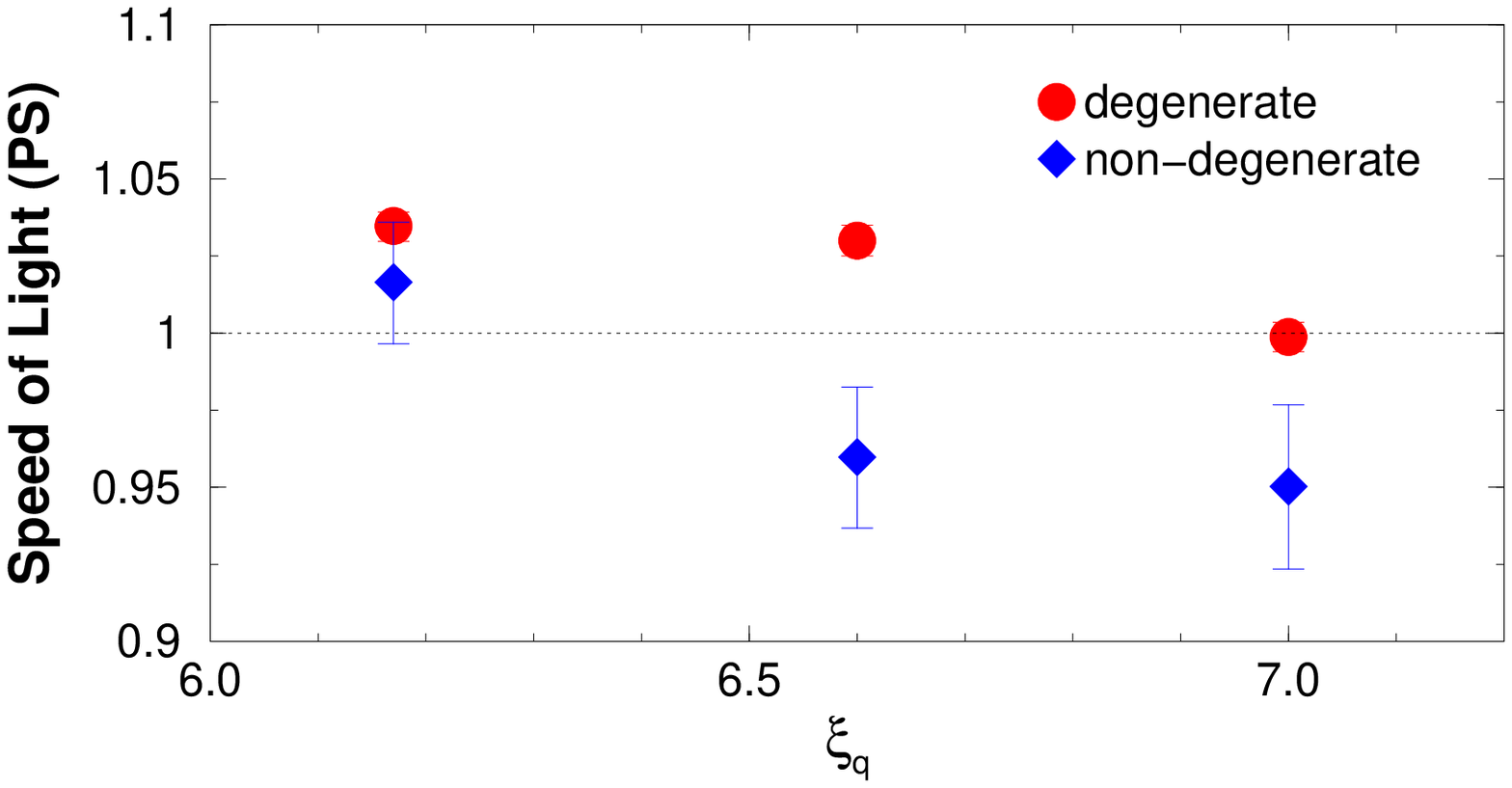}
	\includegraphics[width=7cm]{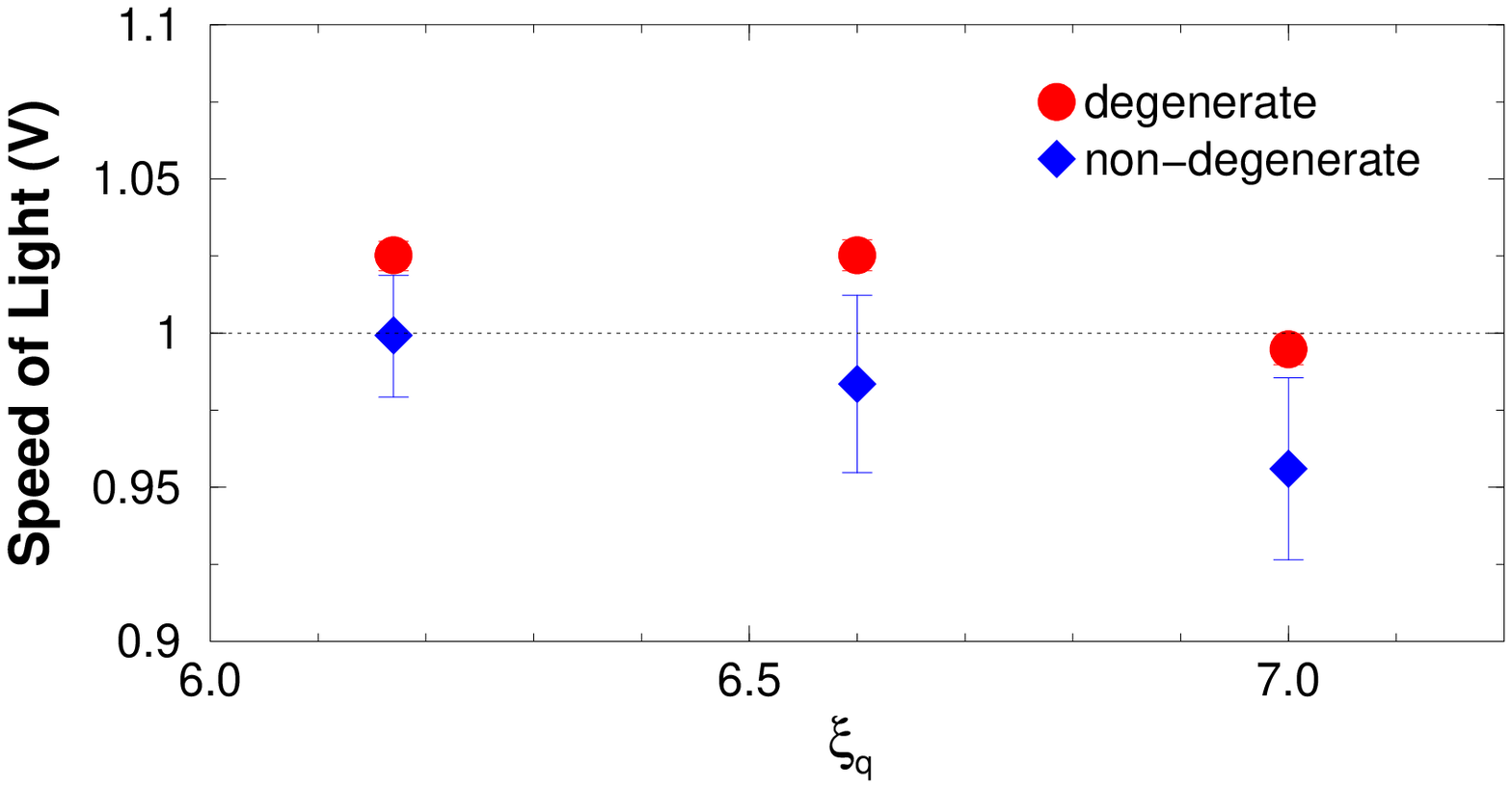}
	\caption[fig:boosted-aniso]{The speed of light as measured from the dispersion
	relation as a function of the parameter, $\xi_q$ in the action. The first plot
	shows the result for the pseudoscalar (PS) meson, for both degenerate and
	non-degenerate combinations. 
	The second plot is the analogous result for the vector (V) mesons. 
	In both cases the quark mass is fixed at $a_tm_q=1.0$. }
	\label{fig:boosted-aniso}
\end{figure}

We have investigated this dependence by varying the parameter $\xi_q$ in the quark action and repeating 
the simulations described above, for the heavy quark mass $a_tm_q=1.0$. 
The dependence of the speed of light, determined from the dispersion relation, on the input 
anisotropy is shown in Figure~\ref{fig:boosted-aniso}. 
%%%%%%%%%%%%%%%%%
%%%%%%%%%%%%%%%%%
The value of $c$ determined from the degenerate meson moves closer to its 
target value of unity and $c$ determined from the non-degenerate physics moves away from this value. 
It is also interesting to note the agreement between determinations of $c$ from pseudoscalar 
and vector particles. The tuning, described above at $a_tm_q=-0.04$ was carried out for pseudoscalars and it 
is reassuring that although the vector particles have larger statistical errors they nevertheless yield 
a consistent picture for the mass dependence of the speed of light.
%%%%%%%%%%%%
\section{Discussion and Conclusions}
\label{sec:conclusions}
In this paper we have explored the viability of anisotropic actions for heavy quark
physics. An action suitable for simulations at large anisotropies and which has no 
${\cal O}(a_sm_q)$ errors is described. One of the main disadvantages of using anisotropic
actions is the extra parameter tuning required to recover Lorentz invariance. In
particular, if the ratio of scales $\xi$ is sensitive to the quark mass in the simulation
then a parameter tuning may be required for each mass. We have measured the speed of light
for a range of quark masses having fixed the ratio of scales at the strange quark,
$a_tm_q=-0.04$. Only slight mass dependence (for the degenerate mesons) is found up, to $a_tm_q = 0.5$ 
which is heavier than the charm quark on these lattices. This implies that one measurement
of the speed of light is all that is required for simulations over a large range of
masses, at the percent-level of simulation. The simulations were repeated for mesons with non-degenerate
quarks, using a value of $\xi$ tuned from the degenerate meson spectrum. The results are
in excellent agreement up to $a_tm_q = 0.5$. Since the charm quark on this lattice is
approximately $a_tm_q=0.2$ this work indicates that both heavy-heavy (degenerate) and
heavy-light (non-degenerate) charm physics can be easily reached
using an appropriately improved anisotropic action. 

The results also show that heavy-light as well as heavy-heavy physics can be reliably
simulated after a single tuning of $\xi_q$. The determination of $c$ can be interpreted as
a measure of the ratio $M_1/M_2$ in Eq. (\ref{eqn:cont-disp}). The agreement of $M_1$ and
$M_2$ for both heavy-heavy and heavy-light systems can in turn be interpreted as an
absence, in this quark action, of the anomaly first discussed in Ref.~\cite{Collins:1996yg}. This anomaly
was explained in Ref.~\cite{Kronfeld:1997uy} where it was pointed out that for a sufficiently accurate
lattice action (${\cal O}(v^4)$ in NRQCD) the discrepancies in binding energies $\delta B
= B_2-B_1$ vanishes and $I=(2\delta B_{\bar{Q}q}-(\delta B_{\bar{Q}Q}+\delta
B_{\bar{q}q}))/2M_{2\bar{Q}q}=0$ as expected. The action described in this study has this property. 

This study has been carried out in the quenched approximation which is a useful laboratory
in which to study mass-dependent and tuning issues at relatively low computational
cost. We are currently developing algorithms for dynamical simulations with anisotropic
lattices which we plan to use in a study of heavy-flavour physics.  
%
%%%%%%%%%%%%%%%%%%%%%
%
\begin{acknowledgments}
The authors would like to thank Jimmy Juge, Colin Morningstar and Jon-Ivar Skullerud for  
carefully reading this manuscript. This work was funded by Enterprise-Ireland grants SC/2001/306 and
SC/2001/307, by IRCSET awards RS/2002/208-7M and SC/2003/393 and under the IITAC PRTLI initiative.
\end{acknowledgments}

\bibliography{AnisoRefs}

\begin{thebibliography}{28}
\expandafter\ifx\csname natexlab\endcsname\relax\def\natexlab#1{#1}\fi
\expandafter\ifx\csname bibnamefont\endcsname\relax
  \def\bibnamefont#1{#1}\fi
\expandafter\ifx\csname bibfnamefont\endcsname\relax
  \def\bibfnamefont#1{#1}\fi
\expandafter\ifx\csname citenamefont\endcsname\relax
  \def\citenamefont#1{#1}\fi
\expandafter\ifx\csname url\endcsname\relax
  \def\url#1{\texttt{#1}}\fi
\expandafter\ifx\csname urlprefix\endcsname\relax\def\urlprefix{URL }\fi
\providecommand{\bibinfo}[2]{#2}
\providecommand{\eprint}[2][]{\url{#2}}

\bibitem[{\citenamefont{Morningstar and Peardon}(1999)}]{Morningstar:1999rf}
\bibinfo{author}{\bibfnamefont{C.~J.} \bibnamefont{Morningstar}}
  \bibnamefont{and} \bibinfo{author}{\bibfnamefont{M.~J.}
  \bibnamefont{Peardon}}, \bibinfo{journal}{Phys. Rev.}
  \textbf{\bibinfo{volume}{D60}}, \bibinfo{pages}{034509}
  (\bibinfo{year}{1999}), \eprint{hep-lat/9901004}.

\bibitem[{\citenamefont{Alford et~al.}(1997)\citenamefont{Alford, Klassen, and
  Lepage}}]{Alford:1997nx}
\bibinfo{author}{\bibfnamefont{M.~G.} \bibnamefont{Alford}},
  \bibinfo{author}{\bibfnamefont{T.~R.} \bibnamefont{Klassen}},
  \bibnamefont{and} \bibinfo{author}{\bibfnamefont{G.~P.}
  \bibnamefont{Lepage}}, \bibinfo{journal}{Nucl. Phys.}
  \textbf{\bibinfo{volume}{B496}}, \bibinfo{pages}{377} (\bibinfo{year}{1997}),
  \eprint{hep-lat/9611010}.

\bibitem[{\citenamefont{Burgio et~al.}(2003{\natexlab{a}})\citenamefont{Burgio,
  Feo, Peardon, and Ryan}}]{Burgio:2003nk}
\bibinfo{author}{\bibfnamefont{G.}~\bibnamefont{Burgio}},
  \bibinfo{author}{\bibfnamefont{A.}~\bibnamefont{Feo}},
  \bibinfo{author}{\bibfnamefont{M.~J.} \bibnamefont{Peardon}},
  \bibnamefont{and} \bibinfo{author}{\bibfnamefont{S.~M.} \bibnamefont{Ryan}}
  (\bibinfo{year}{2003}{\natexlab{a}}), \eprint{hep-lat/0310036}.

\bibitem[{\citenamefont{Burgio et~al.}(2003{\natexlab{b}})\citenamefont{Burgio,
  Feo, Peardon, and Ryan}}]{Burgio:2003gx}
\bibinfo{author}{\bibfnamefont{G.}~\bibnamefont{Burgio}},
  \bibinfo{author}{\bibfnamefont{A.}~\bibnamefont{Feo}},
  \bibinfo{author}{\bibfnamefont{M.~J.} \bibnamefont{Peardon}},
  \bibnamefont{and} \bibinfo{author}{\bibfnamefont{S.~M.} \bibnamefont{Ryan}}
  (\bibinfo{year}{2003}{\natexlab{b}}), \eprint{hep-lat/0309058}.

\bibitem[{\citenamefont{Harada et~al.}(2001)\citenamefont{Harada, Kronfeld,
  Matsufuru, Nakajima, and Onogi}}]{Harada:2001ei}
\bibinfo{author}{\bibfnamefont{J.}~\bibnamefont{Harada}},
  \bibinfo{author}{\bibfnamefont{A.~S.} \bibnamefont{Kronfeld}},
  \bibinfo{author}{\bibfnamefont{H.}~\bibnamefont{Matsufuru}},
  \bibinfo{author}{\bibfnamefont{N.}~\bibnamefont{Nakajima}}, \bibnamefont{and}
  \bibinfo{author}{\bibfnamefont{T.}~\bibnamefont{Onogi}},
  \bibinfo{journal}{Phys. Rev.} \textbf{\bibinfo{volume}{D64}},
  \bibinfo{pages}{074501} (\bibinfo{year}{2001}), \eprint{hep-lat/0103026}.

\bibitem[{\citenamefont{Hashimoto and Okamoto}(2003)}]{Hashimoto:2003fs}
\bibinfo{author}{\bibfnamefont{S.}~\bibnamefont{Hashimoto}} \bibnamefont{and}
  \bibinfo{author}{\bibfnamefont{M.}~\bibnamefont{Okamoto}},
  \bibinfo{journal}{Phys. Rev.} \textbf{\bibinfo{volume}{D67}},
  \bibinfo{pages}{114503} (\bibinfo{year}{2003}), \eprint{hep-lat/0302012}.

\bibitem[{\citenamefont{Foley et~al.}(2003)\citenamefont{Foley, O'Cais,
  Peardon, and Ryan}}]{Foley:2003dh}
\bibinfo{author}{\bibfnamefont{J.}~\bibnamefont{Foley}},
  \bibinfo{author}{\bibfnamefont{A.}~\bibnamefont{O'Cais}},
  \bibinfo{author}{\bibfnamefont{M.~J.} \bibnamefont{Peardon}},
  \bibnamefont{and} \bibinfo{author}{\bibfnamefont{S.~M.} \bibnamefont{Ryan}}
  (\bibinfo{collaboration}{TrinLat}) (\bibinfo{year}{2003}),
  \eprint{hep-lat/0309162}.

\bibitem[{\citenamefont{Hamber and Wu}(1983)}]{Hamber:1983qa}
\bibinfo{author}{\bibfnamefont{H.~W.} \bibnamefont{Hamber}} \bibnamefont{and}
  \bibinfo{author}{\bibfnamefont{C.~M.} \bibnamefont{Wu}},
  \bibinfo{journal}{Phys. Lett.} \textbf{\bibinfo{volume}{B133}},
  \bibinfo{pages}{351} (\bibinfo{year}{1983}).

\bibitem[{\citenamefont{Peardon}(2002)}]{Peardon:2002sd}
\bibinfo{author}{\bibfnamefont{M.}~\bibnamefont{Peardon}},
  \bibinfo{journal}{Nucl. Phys. Proc. Suppl.} \textbf{\bibinfo{volume}{109A}},
  \bibinfo{pages}{212} (\bibinfo{year}{2002}).

\bibitem[{\citenamefont{Okamoto et~al.}(2002)}]{Okamoto:2001jb}
\bibinfo{author}{\bibfnamefont{M.}~\bibnamefont{Okamoto}} \bibnamefont{et~al.}
  (\bibinfo{collaboration}{CP-PACS}), \bibinfo{journal}{Phys. Rev.}
  \textbf{\bibinfo{volume}{D65}}, \bibinfo{pages}{094508}
  (\bibinfo{year}{2002}), \eprint{hep-lat/0112020}.

\bibitem[{\citenamefont{Liao and Manke}(2002)}]{Liao:2001yh}
\bibinfo{author}{\bibfnamefont{X.}~\bibnamefont{Liao}} \bibnamefont{and}
  \bibinfo{author}{\bibfnamefont{T.}~\bibnamefont{Manke}},
  \bibinfo{journal}{Phys. Rev.} \textbf{\bibinfo{volume}{D65}},
  \bibinfo{pages}{074508} (\bibinfo{year}{2002}), \eprint{hep-lat/0111049}.

\bibitem[{\citenamefont{Chen}(2001)}]{Chen:2000ej}
\bibinfo{author}{\bibfnamefont{P.}~\bibnamefont{Chen}}, \bibinfo{journal}{Phys.
  Rev.} \textbf{\bibinfo{volume}{D64}}, \bibinfo{pages}{034509}
  (\bibinfo{year}{2001}), \eprint{hep-lat/0006019}.

\bibitem[{\citenamefont{Edwards et~al.}(2003)\citenamefont{Edwards, Heller, and
  Richards}}]{Edwards:2003cd}
\bibinfo{author}{\bibfnamefont{R.~G.} \bibnamefont{Edwards}},
  \bibinfo{author}{\bibfnamefont{U.~M.} \bibnamefont{Heller}},
  \bibnamefont{and} \bibinfo{author}{\bibfnamefont{D.~G.}
  \bibnamefont{Richards}} (\bibinfo{collaboration}{LHP}),
  \bibinfo{journal}{Nucl. Phys. Proc. Suppl.} \textbf{\bibinfo{volume}{119}},
  \bibinfo{pages}{305} (\bibinfo{year}{2003}), \eprint{hep-lat/0303004}.

\bibitem[{\citenamefont{Matsufuru et~al.}(2003)\citenamefont{Matsufuru, Harada,
  Onogi, and Sugita}}]{Matsufuru:2002vh}
\bibinfo{author}{\bibfnamefont{H.}~\bibnamefont{Matsufuru}},
  \bibinfo{author}{\bibfnamefont{J.}~\bibnamefont{Harada}},
  \bibinfo{author}{\bibfnamefont{T.}~\bibnamefont{Onogi}}, \bibnamefont{and}
  \bibinfo{author}{\bibfnamefont{A.}~\bibnamefont{Sugita}},
  \bibinfo{journal}{Nucl. Phys. Proc. Suppl.} \textbf{\bibinfo{volume}{119}},
  \bibinfo{pages}{601} (\bibinfo{year}{2003}), \eprint{hep-lat/0209090}.

\bibitem[{\citenamefont{Luo and Mei}(2003)}]{Luo:2002rz}
\bibinfo{author}{\bibfnamefont{X.-Q.} \bibnamefont{Luo}} \bibnamefont{and}
  \bibinfo{author}{\bibfnamefont{Z.-H.} \bibnamefont{Mei}},
  \bibinfo{journal}{Nucl. Phys. Proc. Suppl.} \textbf{\bibinfo{volume}{119}},
  \bibinfo{pages}{263} (\bibinfo{year}{2003}), \eprint{hep-lat/0209049}.

\bibitem[{\citenamefont{Harada et~al.}(2002)\citenamefont{Harada, Matsufuru,
  Onogi, and Sugita}}]{Harada:2002ii}
\bibinfo{author}{\bibfnamefont{J.}~\bibnamefont{Harada}},
  \bibinfo{author}{\bibfnamefont{H.}~\bibnamefont{Matsufuru}},
  \bibinfo{author}{\bibfnamefont{T.}~\bibnamefont{Onogi}}, \bibnamefont{and}
  \bibinfo{author}{\bibfnamefont{A.}~\bibnamefont{Sugita}},
  \bibinfo{journal}{Nucl. Phys. Proc. Suppl.} \textbf{\bibinfo{volume}{111}},
  \bibinfo{pages}{282} (\bibinfo{year}{2002}), \eprint{hep-lat/0202004}.

\bibitem[{\citenamefont{Shigemitsu et~al.}(2002)}]{Shigemitsu:2002wh}
\bibinfo{author}{\bibfnamefont{J.}~\bibnamefont{Shigemitsu}}
  \bibnamefont{et~al.}, \bibinfo{journal}{Phys. Rev.}
  \textbf{\bibinfo{volume}{D66}}, \bibinfo{pages}{074506}
  (\bibinfo{year}{2002}), \eprint{hep-lat/0207011}.

\bibitem[{\citenamefont{de~Forcrand et~al.}(2000)}]{deForcrand:1999df}
\bibinfo{author}{\bibfnamefont{P.}~\bibnamefont{de~Forcrand}}
  \bibnamefont{et~al.} (\bibinfo{collaboration}{QCD-TARO}),
  \bibinfo{journal}{Nucl. Phys. Proc. Suppl.} \textbf{\bibinfo{volume}{83}},
  \bibinfo{pages}{411} (\bibinfo{year}{2000}), \eprint{hep-lat/9911001}.

\bibitem[{\citenamefont{Umeda et~al.}(2001)\citenamefont{Umeda, Katayama,
  Miyamura, and Matsufuru}}]{Umeda:2000ym}
\bibinfo{author}{\bibfnamefont{T.}~\bibnamefont{Umeda}},
  \bibinfo{author}{\bibfnamefont{R.}~\bibnamefont{Katayama}},
  \bibinfo{author}{\bibfnamefont{O.}~\bibnamefont{Miyamura}}, \bibnamefont{and}
  \bibinfo{author}{\bibfnamefont{H.}~\bibnamefont{Matsufuru}},
  \bibinfo{journal}{Int. J. Mod. Phys.} \textbf{\bibinfo{volume}{A16}},
  \bibinfo{pages}{2215} (\bibinfo{year}{2001}), \eprint{hep-lat/0011085}.

\bibitem[{\citenamefont{Klassen}(1998)}]{Klassen:1998jf}
\bibinfo{author}{\bibfnamefont{T.~R.} \bibnamefont{Klassen}},
  \bibinfo{journal}{Nucl. Phys.} \textbf{\bibinfo{volume}{B509}},
  \bibinfo{pages}{391} (\bibinfo{year}{1998}), \eprint{hep-lat/9705025}.

\bibitem[{\citenamefont{Klassen}(1999)}]{Klassen:1998fh}
\bibinfo{author}{\bibfnamefont{T.~R.} \bibnamefont{Klassen}},
  \bibinfo{journal}{Nucl. Phys. Proc. Suppl.} \textbf{\bibinfo{volume}{73}},
  \bibinfo{pages}{918} (\bibinfo{year}{1999}), \eprint{hep-lat/9809174}.

\bibitem[{\citenamefont{Chen et~al.}(2001)\citenamefont{Chen, Liao, and
  Manke}}]{Chen:2000qj}
\bibinfo{author}{\bibfnamefont{P.}~\bibnamefont{Chen}},
  \bibinfo{author}{\bibfnamefont{X.}~\bibnamefont{Liao}}, \bibnamefont{and}
  \bibinfo{author}{\bibfnamefont{T.}~\bibnamefont{Manke}},
  \bibinfo{journal}{Nucl. Phys. Proc. Suppl.} \textbf{\bibinfo{volume}{94}},
  \bibinfo{pages}{342} (\bibinfo{year}{2001}), \eprint{hep-lat/0010069}.

\bibitem[{\citenamefont{Ali~Khan et~al.}(2001)}]{AliKhan:2000bv}
\bibinfo{author}{\bibfnamefont{A.}~\bibnamefont{Ali~Khan}} \bibnamefont{et~al.}
  (\bibinfo{collaboration}{CP-PACS}), \bibinfo{journal}{Nucl. Phys. Proc.
  Suppl.} \textbf{\bibinfo{volume}{94}}, \bibinfo{pages}{325}
  (\bibinfo{year}{2001}), \eprint{hep-lat/0011005}.

\bibitem[{\citenamefont{El-Khadra et~al.}(1997)\citenamefont{El-Khadra,
  Kronfeld, and Mackenzie}}]{El-Khadra:1997mp}
\bibinfo{author}{\bibfnamefont{A.~X.} \bibnamefont{El-Khadra}},
  \bibinfo{author}{\bibfnamefont{A.~S.} \bibnamefont{Kronfeld}},
  \bibnamefont{and} \bibinfo{author}{\bibfnamefont{P.~B.}
  \bibnamefont{Mackenzie}}, \bibinfo{journal}{Phys. Rev.}
  \textbf{\bibinfo{volume}{D55}}, \bibinfo{pages}{3933} (\bibinfo{year}{1997}),
  \eprint{hep-lat/9604004}.

\bibitem[{\citenamefont{Morningstar and Peardon}(2000)}]{Morningstar:1999dh}
\bibinfo{author}{\bibfnamefont{C.}~\bibnamefont{Morningstar}} \bibnamefont{and}
  \bibinfo{author}{\bibfnamefont{M.~J.} \bibnamefont{Peardon}},
  \bibinfo{journal}{Nucl. Phys. Proc. Suppl.} \textbf{\bibinfo{volume}{83}},
  \bibinfo{pages}{887} (\bibinfo{year}{2000}), \eprint{hep-lat/9911003}.

\bibitem[{\citenamefont{Morningstar and Peardon}(2004)}]{Morningstar:2003gk}
\bibinfo{author}{\bibfnamefont{C.}~\bibnamefont{Morningstar}} \bibnamefont{and}
  \bibinfo{author}{\bibfnamefont{M.~J.} \bibnamefont{Peardon}},
  \bibinfo{journal}{Phys. Rev.} \textbf{\bibinfo{volume}{D69}},
  \bibinfo{pages}{054501} (\bibinfo{year}{2004}), \eprint{hep-lat/0311018}.

\bibitem[{\citenamefont{Collins et~al.}(1996)\citenamefont{Collins, Edwards,
  Heller, and Sloan}}]{Collins:1996yg}
\bibinfo{author}{\bibfnamefont{S.}~\bibnamefont{Collins}},
  \bibinfo{author}{\bibfnamefont{R.~G.} \bibnamefont{Edwards}},
  \bibinfo{author}{\bibfnamefont{U.~M.} \bibnamefont{Heller}},
  \bibnamefont{and} \bibinfo{author}{\bibfnamefont{J.~H.} \bibnamefont{Sloan}},
  \bibinfo{journal}{Nucl. Phys. Proc. Suppl.} \textbf{\bibinfo{volume}{47}},
  \bibinfo{pages}{455} (\bibinfo{year}{1996}), \eprint{hep-lat/9512026}.

\bibitem[{\citenamefont{Kronfeld}(1997)}]{Kronfeld:1997uy}
\bibinfo{author}{\bibfnamefont{A.~S.} \bibnamefont{Kronfeld}},
  \bibinfo{journal}{Nucl. Phys. Proc. Suppl.} \textbf{\bibinfo{volume}{53}},
  \bibinfo{pages}{401} (\bibinfo{year}{1997}), \eprint{hep-lat/9608139}.

\end{thebibliography}

\end{document}